\documentclass[journal=jacsat,manuscript=article]{achemso}

\usepackage{chemformula} 
\usepackage[T1]{fontenc} 

\author{Toru Matsuura}
\email{t-matsuura@fukui-nct.ac.jp}
\affiliation
{Department of Electrical and Electronic Engineering, National Institute of Technology (KOSEN), Fukui College, Geshi-cho, Sabae-shi, Fukui 916-8507, Japan}
\author{Kazuma Handa}
\affiliation
{Production System Engineering Course, National Institute of Technology (KOSEN), Fukui College, Geshi-cho, Sabae-shi, Fukui 916-8507, Japan}
\author{Masakazu Arakawa}
\affiliation
{Department of Electrical and Electronic Engineering, National Institute of Technology (KOSEN), Fukui College, Geshi-cho, Sabae-shi, Fukui 916-8507, Japan}

\title
{ Efficiency of negative-illumination photovoltaic energy conversion}

\begin{document}


\begin{abstract}
Infrared diodes generate electricity from thermal radiation emitted from themselves.
The negative process of photovoltaic effect has been expected for application to energy harvesting systems converting from terrestrial radiation.
However, its energy conversion efficiency has been known to be very low.
In this paper, we investigate energy conversion efficiency and external quantum efficiency for the negative-illumination photovoltaic effect with a systematic measurement for infrared diodes faced to a sufficiently cold surface. 
We find that the external quantum efficiency reaches 60\ \% for a diode at a temperature, while the energy conversion efficiency stays below $10^{-4}\ \%$.
We indicate dominant parameters for the efficiencies and propose how to improve energy conversion efficiency.
\end{abstract}

\section{Introduction}
The large temperature difference between the cold outer space and the hot Earth's surface has a potential to operating a heat engine \cite{Byrnes2014PNAS}. 
The outer space is filled by microwave radiation everywhere from any direction due to remnants of the Big Bang\cite{Penzias1965AstroJ}.
The microwave spectrum corresponds to that of the blackbody radiation at 2.725K.
Thus, outer space is regarded as an enormous cold heat bath.
In contrast, Earth always emits infrared light into outer space.
The thermal radiation, known as terrestrial radiation, has a power density of 400 W/m$^2$, and plays an important role for maintaining the average global temperature on Earth.
The energy utilization of the terrestrial radiation has been investigated using several approaches, such as a radiative cooling refrigerator \cite{Bhatia2018Naturecomm}, a radiative thermoelectric generator \cite{Assawaworrarit2022APL}, water harvesting \cite{Minghao2021ACS}, and direct conversion to electricity, called the negative-illumination photovoltaic effect or thermoradiative photovoltaic effect with a p-n junction \cite{Santhanam2016PRB,Ono2019APL,Nielsen2022ACS}, junction-free strained Bi$_2$Te$_3$ thin films having the bulk-photovoltaic effect \cite{Lorenzi2020OpEx}, and infrared rectenna arrays \cite{Belkadi2023ACS}.
Furthermore, indirect negative-illumination photovoltaic effect using a heat absorber \cite{Liao2017OptLett}, and the hybrid systems with the thermoelectric effect and the negative-illumination photovoltaic effect \cite{Lorenzi2022ACS} have been reported. 

A p-n junction acts as a heat engine to convert electrical energy $W$ from a temperature deference, as shown in fig. \ref{fig1} (a).
The thermal radiation from the p-n junction to the cold bath carries heat $Q_L$, and it induces heat flux $Q_H$ from the hot bath to the p-n junction.
Since thermal radiation involves carrier recombination processes, imbalances in minority carrier densities across the p-n junction makes negative photocurrent.
The process is a reverse process of the conventional photovoltaic effect, and the electrical energy is converted from a part of the heat flux $Q_H$. 
The energy conversion efficiency of this engine is given by 
\begin{eqnarray} 
\eta_E = \frac{W}{Q_H} \leq 1-\frac{T_L}{T_H}.
\label{efficiency}
\end{eqnarray}
The Carnot efficiency $\eta_C=1-\frac{T_L}{T_H}$ is the upper bound of efficiency. 
For the case of $T_L = 3\ \mathrm{K}$ and $T_H = 300\ \mathrm{K}$, the efficiency can reach 99\ \% in principle.

The negative-illumination photovoltaic effect has been reported with p-n junction infrared diodes based on HgCdTe \cite{Santhanam2016PRB,Ono2019APL,Nielsen2022ACS}.
In the most recent demonstration\cite{Nielsen2022ACS}, the output power was achieved to $2.26\ \mathrm{mW/m^2}$ for the temperature difference of $12.5\ \mathrm{K}$.
Their expectation for an ideal condition using the outer space as the cold bath estimated the maximum output power could reach $19.4\ \mathrm{mW/m^2}$ for a mid-infrared diode of 0.2 eV bandgap.
However, it was still only $0.005 \%$ of the terrestrial radiation.

An indoor experiment with a thermoelectric cooler (TEC) for a cold surface could avoid the atmosphere absorption but a TEC could not reach a sufficient temperature difference to observe the maximum output power of the diodes \cite{Nielsen2022ACS}.
In contrast, the previous experimental approaches using the real night sky \cite{Santhanam2016PRB,Ono2019APL}, the effect of the atmosphere absorption could not be excluded.
The potential efficiency with and without the atmosphere absorption has been under discussion \cite{Stanfberg2015JAP,Lin2017JAP,Zhang2020OptLett,Callahan2021PRApp,Shibuya2022JTST,Buddhiraju2018PNAS,Deppe2020ACS,Harrison2024IEEE,Nielsen2024NaturePhotonics}.
However, the maximum output power under negative-illumination that p-n junctions possess inherently has not been observed directly yet.

In this paper, we investigate inherent ability of electrical power generation using infrared p-n junction diodes under the negative-illumination condition without atmosphere absorption as a function of the temperature of the cold bath and that of the hot bath.
We analyze the temperature dependence of the photovoltage with the single homo junction model and estimate the external quantum efficiency and the energy conversion efficiency.
The dominant physical factors for the efficiencies are discussed, which have never been treated explicitly in the previous theoretical approaches.

\section{Experimental}
The measurement is performed with our experimental setup shown in Figure \ref{fig1} (a).
An infrared diode mounted on the hot bath stage of copper block.
The diode is faced with a cold bath of copper block cooled down by liquid nitrogen.
The separating distance is 50 mm.
The surface of the cold bath is coated by the blackbody painting to maximize its optical absorbance.
The metal reflector is put on the hot bath to lead thermal radiation emitted from the diode to the cold bath.
The temperature of the hot bath $T_H$ and that of cold bath $T_L$ are measured by Pt100 resistance thermometers.
The hot bath temperature is controlled by an electric heater.
The diodes and the baths are set in a vacuum chamber to avoid frosting on the cold bath and air convection.
The voltmeter (Keithley 2000 or Keithley 2182A) measures the diode photovoltage, and the dc current source (Keithley 6220) applies current for I-V characteristic measurement.

Two p-n junction devices consisting of a narrow direct bandgap semiconductor \cite{Chiang1985APL,Murawski2019PNSMI} are used for the experiment.
The first one is an indium arsenide antimonide (InAsSb) far-infrared photovoltaic device (Hamamatsu Photonics, P13894-011NA) with the bandgap energy $E_g = 0.12\ \mathrm{eV}$, and the second one is a mid-infrared diode (Hamamatsu Photonics, L13201-0430M).
The former is called Diode A, and the latter is called Diode B.
The bandgap energy $E_g$ of Diode A is $0.12\ \mathrm{eV}$, and that of Diode B is $0.26\ \mathrm{eV}$.
Since the maximum of the energy flux density of thermal radiation from the blackbody at 300 K is at a photon energy of $0.124\ \mathrm{eV}$, Diode A seems to be appropriate for energy conversion from terrestrial radiation rather than Diode B. 

The negative-illumination photovoltaic effect must be associated with only spontaneous radiative recombination across the bandgap for InAsSb, and not with contribution of phonon.
The optical phonon blanches would appear at about $200\ \mathrm{cm}^{-1}$ in InAsSb \cite{Keifer1975PRB,Carles1980PRB}, and the radiation from the optical phonon relaxation should have 5 to 6 times longer wavelength than $10\ \mu\mathrm{m}$. 

Each diode device is configured by a series circuit of several small p-n junctions connected by metallic electrodes on a tip.
The number $N$ of single diodes for Diode A is about 300, and that for Diode B is 7, respectively. 
Figure \ref{fig2} (b) shows the equivalent circuit of a series connection of p-n junctions.
The current sources represent photocurrent generation, and the parallel resistances are shunt resistances of the p-n junctions.
The series resistance is neglected.
The voltage of the device $V_{D}$ is the sum of the voltages in the series p-n junctions, such as $V_{D} = N\times v_{D}$.
Similarly, the zero-bias resistance $R_D$ is the sum of the resistances in the series p-n junctions, such as $R_{D} = N\times r_{D}$.
The photocurrent is given by short-circuit current $I_{SC}$, and it flows commonly in each single p-n junction.
Additionally, the light emitting area of a single p-n junction $S$ is measured as $600\ \mathrm{\mu m^2}$ for Diode A, and as $50000\ \mathrm{\mu m^2}$ for Diode B from microscope observation.
Since the top surface of the devices are partially covered by some metallic electrodes and pathways, only unshaded areas can contribute to the photovoltaic effect.

\section{Results}
\begin{figure*}[t]
\centering
\includegraphics[width=0.8\linewidth]{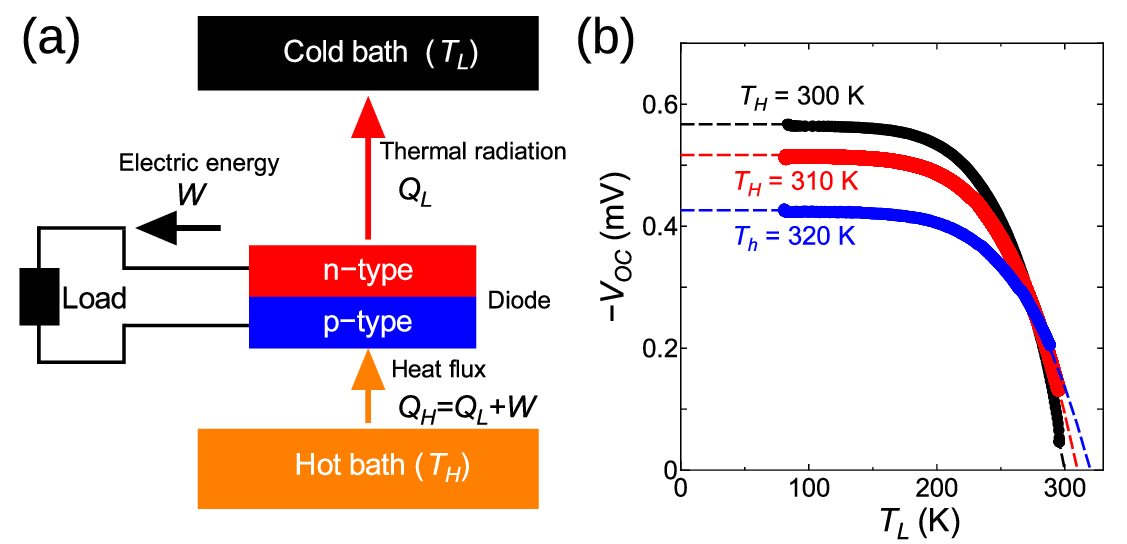} 
\vspace{0cm}
\caption{(a) Negative-illumination photovoltaic generator as a heat engine. 
(b) Photovoltage for an InAsSb photovoltaic cell (Diode A) under negative-illumination as a fuction of cold bath temperature.
The thin dashed curves are theoretical curves.}
\label{fig1}
\end{figure*}

\begin{figure*}[t]
\centering
\includegraphics[width=0.8\linewidth]{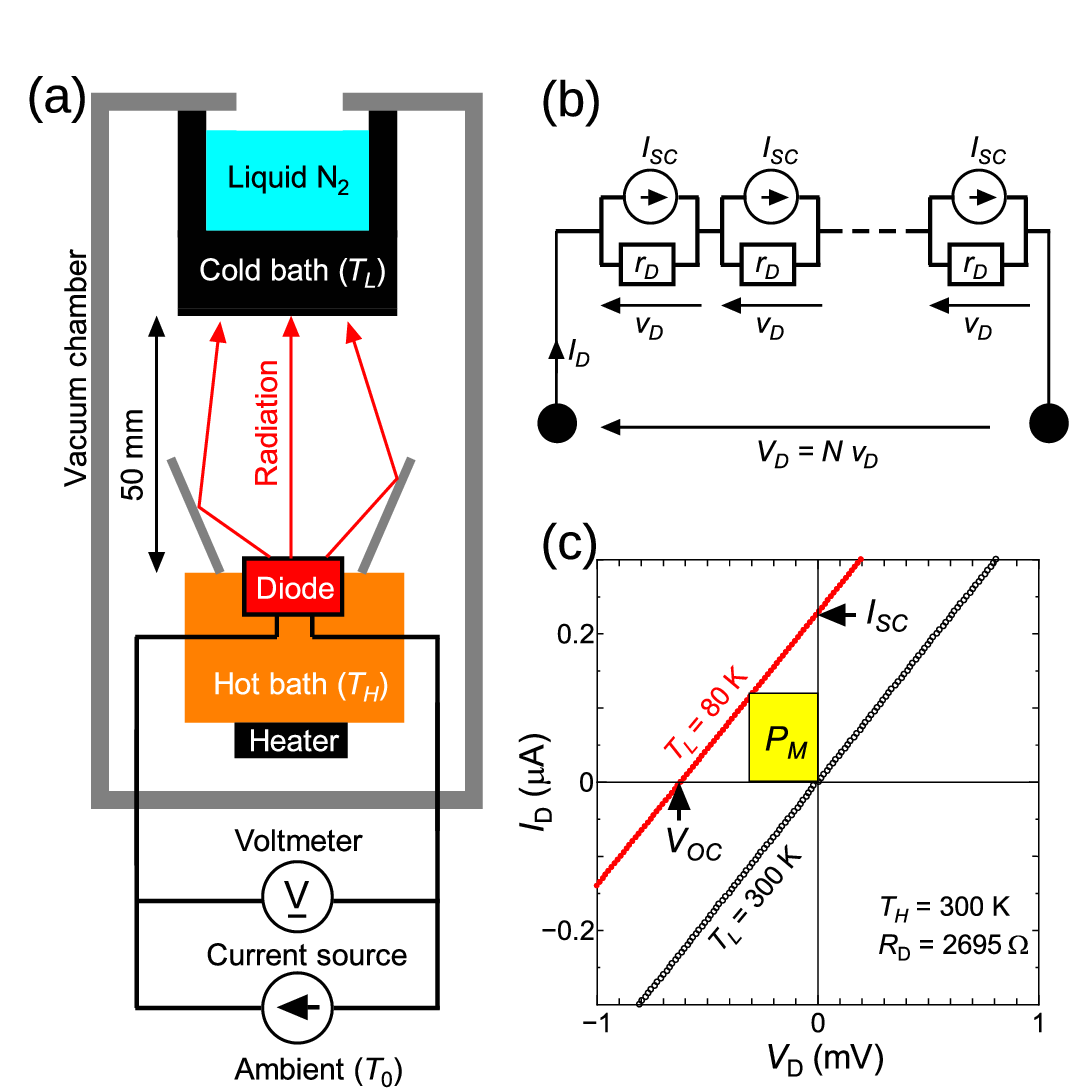} 
\vspace{0cm}
\caption{(a) Schematics of experimental setup. 
(b) Equivalent circuit of a diode device configured by a series circuit of several single p-n junctions. 
(c) Current voltage characteristics of Diode A for $T_L = 300\ \mathrm{K}$ (in equilibrium) and $80\ \mathrm{K}$ (under negative-illumination).
The area of square $P_M$ corresponds to the maximum power generation under negative-illumination.}
\label{fig2}
\end{figure*}

Figure \ref{fig1} (b) shows the experimental results of the open-circuit voltage of Diode A as a function of the cold bath temperature $T_L$.
The photovoltage is obtained by the open-circuit voltage with negative sign, as shown in Fig. \ref{fig2} (c).
The hot bath temperature $T_H$ is kept constant by electric heater control, and the diode temperature is identified with the hot bath temperature $T_H$, since the heat conduction between the hot bath and the diode is well, 
As $T_L$ decreases, the photovoltage $-V_{OC}$ increases, and then saturates to be constant. 
This behavior indicates that thermal radiation from the cold bath at the liquid nitrogen temperature (77 K) is negligible.
Thus, the cold bath is sufficiently cold to substitute the outer space.

Furthermore. we find that the saturated voltages for $T_H = 300\ \mathrm{K}$ is larger than that for $T_H = 320\ \mathrm{K}$.
Generally, thermal radiation of a hotter object is expected to be larger than that of a colder object.
The experimental result is contrary to the expectation.

\begin{figure}[t]
\centering
\includegraphics[width=0.8\linewidth]{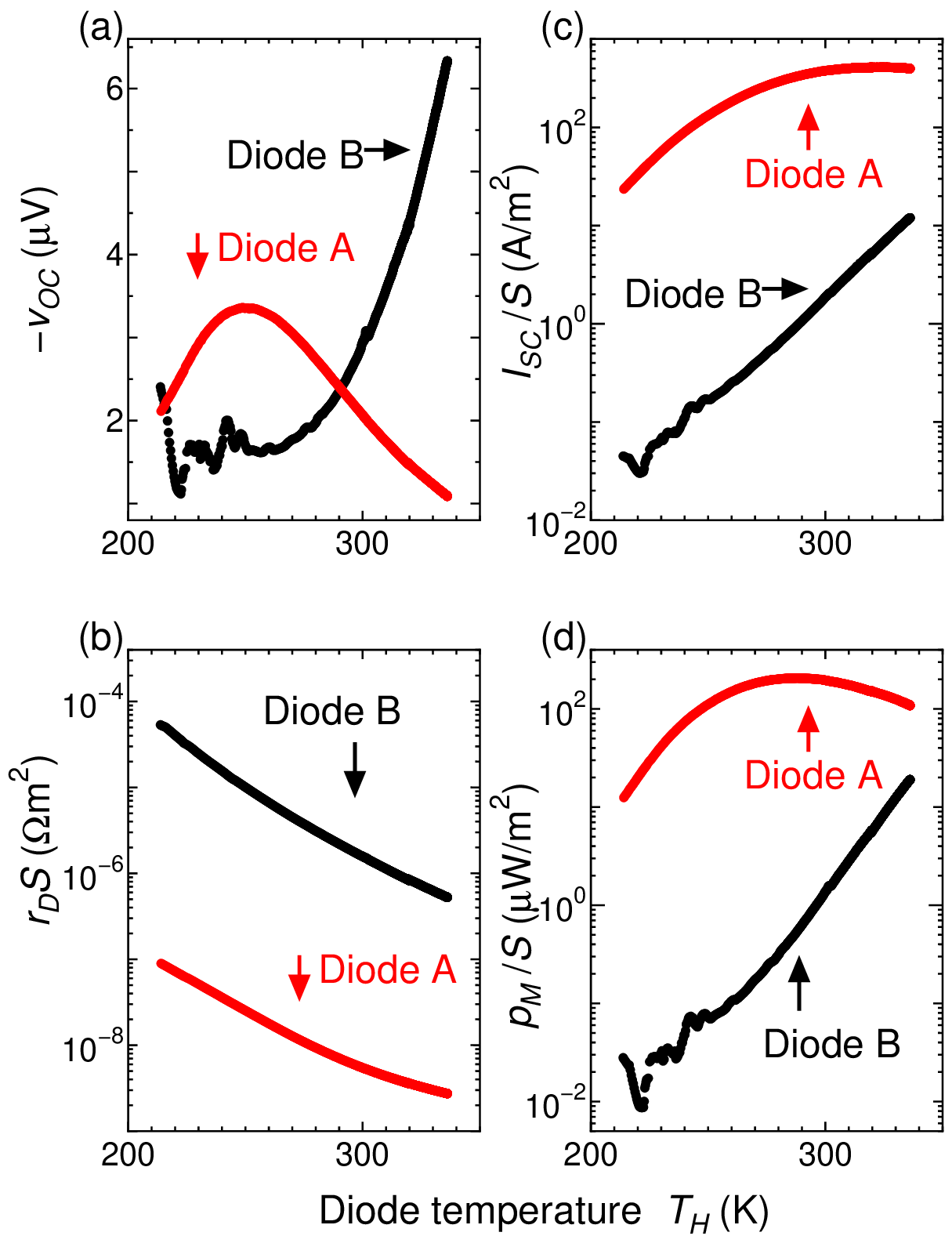} 
\vspace{2cm}
\caption{Diode temperature dependence of photovoltage of a single p-n junction $-v_{OC} = -V_{OC}/N$ (a), zero-bias resistance of a single p-n junction for unit area $r_D S$ (b), inherent photocurrent density $I_{SC}/S$, and maximum power generation density $p_{M}/S$ (d). The cold bath temperature $T_L$ is kept in the range between $80$ and $85\ \mathrm{K}$. }
\label{fig3}
\end{figure}

The photovoltage of a single p-n junction $-v_{OC}=-V_{OC}/N$ for Diode A and B as a function of $T_H$ is shown in Fig. \ref{fig3} (a). 
The cold bath temperature $T_L$ is kept at $80 \sim 85 \ \mathrm{K}$.
The hot bath and diode is cooled by heat exchange gas (helium) introduced in the vacuum chamber. 
The measurement is performed after evacuating the gas, and $T_H$ is controlled by an electrical heater on the hot stage.
The curves of $-v_{OC}$ in Fig. \ref{fig3} (a) has been corrected with thermal electromotive force. 
In this temperature sweeping measurement, the thermal electromotive force caused by temperature difference between the diode at $T_H$ and the voltmeter at $T_0$ is added to inherent photovoltage.
In the previous experiments, an optical chopper was often used to avoid this problem \cite{Santhanam2016PRB,Ono2019APL,Nielsen2022ACS}.
However, the chopper mechanism could not install it in our vacuum chamber.
Thus, the thermal electromotive force is measured while the diode is shaded by a metal cover to correct the photovoltage.

Notably, $-v_{OC}$ of Diode A has the maximum value at 250 K.
Above 250 K, $-v_{OC}$ decreases as the diode temperature increases, which is consistent with the observation in Fig. \ref{fig1} (b).
Since $-v_{OC}$ increases as the diode temperature increases below 250 K, it seems to be zero when the diode temperature equals the temperature of the cold bath.
Therefore, it is no violation of thermodynamics. 
In contrast, $-v_{OC}$ of Diode B shows a monotonic increase as $T_H$ increases. 

As shown in Fig. \ref{fig2} (c), both diodes indicate linear I-V characteristics in the voltage range of our experiment.
For equilibrium, the linear curve passes through the origin.
On the other hand, under the negative-illumination, the linear curve moves up due to the photocurrent generation. 
Thus, 
\begin{eqnarray} 
I_{D} = V_{D}/R_D + I_{SC}.
\label{I_D}
\end{eqnarray}
The photocurrent is given by short-circuit current $I_{SC}$.
The maximum electric power generation $P_{M}$ are given by 
\begin{eqnarray} 
P_{M} = -V_{OC} I_{SC}/4 > 0.
\label{I_ph}
\end{eqnarray}
The factor 1/4 of $P_{M}$ is introduced by the maximum power transfer theorem.

Figures \ref{fig3} (b)$\sim$ \ref{fig3} (d) show zero-bias resistance, inherent photocurrent density, and maximum electrical power generation density of a single p-n junction, respectively.
The peak value of the maximum power generation of Diode A is $206\ \mathrm{\mu W/m^2}$ at 288 K, which is same order of the previous researches in HgTe based diodes \cite{Santhanam2016PRB,Ono2019APL,Lorenzi2020OpEx,Nielsen2022ACS}. 

\section{Discussions}
The negative-illumination photovoltage ($V_{OC}<0$) is expressed by 
\begin{eqnarray} 
-V_{OC}(T_H, T_L) &=& I_{SC}(T_H, T_L) R_D(T_H) \nonumber \\
&=& -q \eta_Q(T_H) \left[\rho(T_L) - \rho(T_H) \right] S R_D(T_H),
\label{V_ph_fit}
\end{eqnarray}
where, $q$ is the elementary charge.
$\rho(T_L)$ and $\rho(T_H)$ are photon flux density emitted from the cold surface and that emitted from the diode surface. 
The photocurrent $I_{SC}$ is proportional to difference between number of emitting photons and that of incident photons.
Using the Planck law, radiation spectrum emitting from a semiconductor with bandgap energy $E_g$ can be expressed as 
\begin{eqnarray} 
\rho(T) = \int_{E_g/h}^\infty \frac{2\pi\nu^2 }{c^2}\frac{1}{\exp\{h\nu/k_BT\}-1} d\nu,
\label{PFD}
\end{eqnarray} 
where $c$, $h$ and $k_B$ are the speed of light in vacuum, the Planck constant and the Boltzmann constant, respectively.

$\eta_Q(T_H)$ is defined as the external quantum efficiency, which means the ratio between the number of charges contributing to the photocurrent and the number of photons radiating from the diode surface in the negative illumination condition.
The theoretical curves in Fig. \ref{fig1} (b) fit well to the experimental curves in the whole measurement temperature range with 
only one fitting parameter; $\eta_Q(300\ \mathrm{K}) = 40.8\ \%$, $\eta_Q(310\ \mathrm{K}) = 38.0\ \%$, and $\eta_Q(330\ \mathrm{K}) = 31.6\ \%$, respectively.
The results show that the temperature dependence of $\eta_Q$ plays an important role for the strange temperature dependence of $V_{OC}$.

The energy conversion efficiency $\eta_E$ in eq. (\ref{efficiency}) for the negative-illumination photovoltaic effect depends on $\eta_Q$. 
Heat moves from the diode to the cold bath is assumed to be the blackbody radiation, then $Q_L = \int\sigma T_H^4 S dt$, where $\sigma T_H^4$ is the radiation energy density given by the Stefan-Boltzmann law and $\sigma = \frac{2\pi^5k_B^4}{15c^2h^3}$ is the Stefan-Boltzmann constant. 
The maximum work that the diode generates is given by $W = \int p_{M} dt$. 
Heat from the hot bath to the diode is $Q_H = Q_L +W$.
Then substituting eq. (\ref{V_ph_fit}) and eq. (\ref{PFD}) in eq. (\ref{efficiency}), we obtain the following expression,
\begin{eqnarray} 
\eta_E &=&\frac{p_{M}}{\sigma T_H^4S+ p_{M}}\nonumber \\
&=& \frac{\zeta \eta^2y(x_g)^2T_H^2r_DS}{1+\zeta \eta^2y(x_g)^2T_H^2r_DS},
\label{chi}
\end{eqnarray}
where $\zeta = \frac{15q^2k_B^2}{2\pi^3c^3h^3}$, $y(x_g) = \int^\infty_{x_g} \frac{x^2dx}{e^x-1}$, and $x_g = E_g/k_BT_H$. 
Here, radiation photon flux density from the cold surface $\rho(T_L)$ is assumed to be zero.
Thus, the energy conversion efficiency of eq. (\ref{chi}) is the maximum efficiency inherently depend on only the diode.

\begin{figure}[t]
\centering
\includegraphics[width=0.8\linewidth]{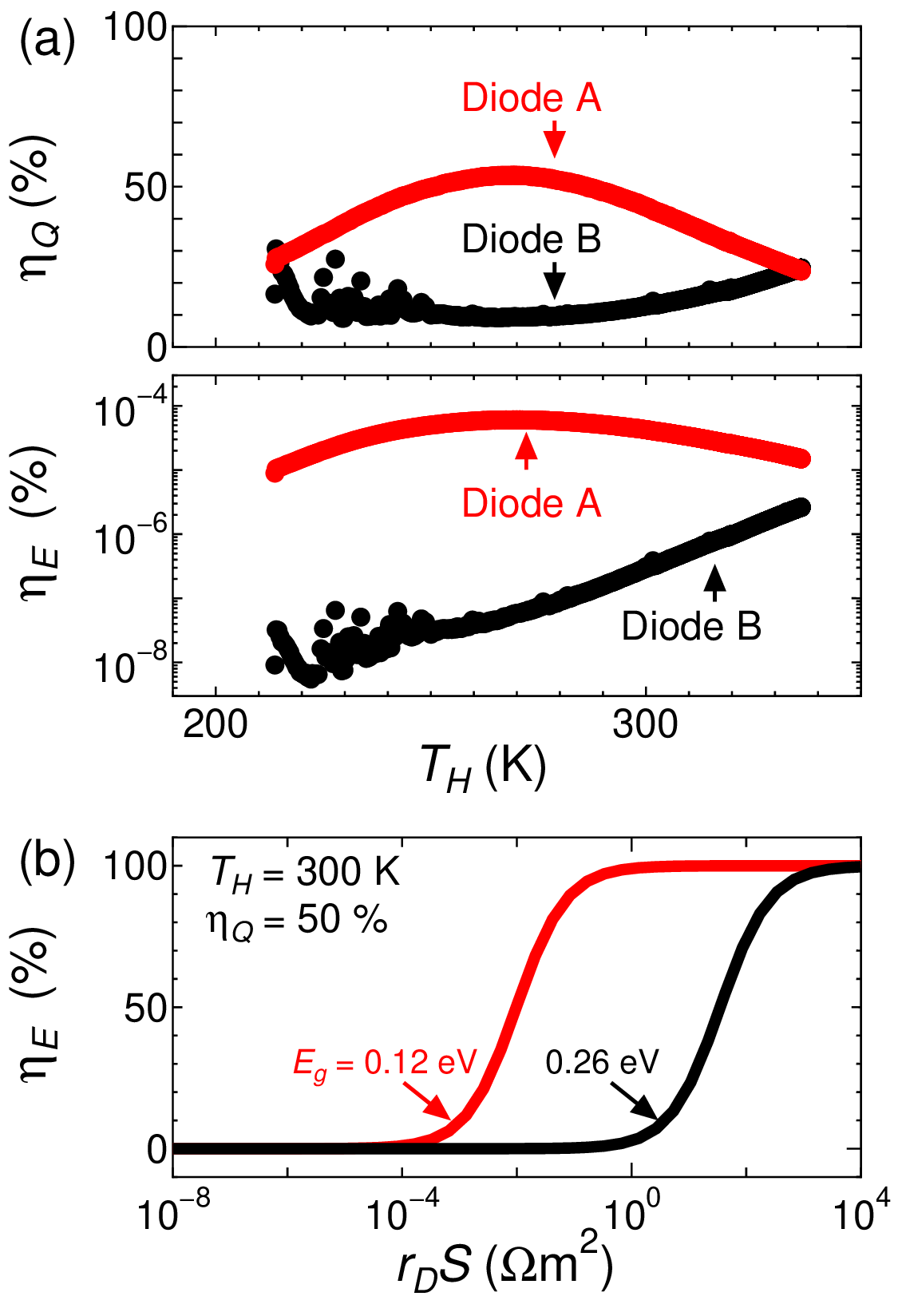} 
\vspace{2cm}
\caption{(a) External quantum efficiency $\eta_Q$ and energy conversion efficiency $\eta_E$ as a function of inverse diode temperature. 
(b) Theoretical prediction of $\eta_E$ as a function of $r_D S$.
}
\label{fig4}
\end{figure}

The experimental results of the external quantum efficiency $\eta_Q(T_H)$ and the energy conversion efficiency $\eta_E(T_H)$ are shown in Figure \ref{fig4} (a).
For Diode A, $\eta_Q$ is in the range between 10 to 60\ \% in the measured temperature range.
For Diode B, $\eta_Q$ is lower than that of Diode A. 
The order of magnitude of $\eta_Q$ for two diodes indicates that the diode devices show substantially good performance.
The temperature dependence of $\eta_Q$ for Diode A is different from that of Diode B. 
For Diode A, $\eta_Q$ increases as $T_H$ increases, and then it shows maximum at 270 K. 
Above 270 K, $\eta_Q$ decreases as $T_H$ increases, which corresponds to the result in Figure \ref{fig1} (b).
In contrast, $\eta_Q$ of Diode B shows a monotonic increase as temperature increases.

The temperature dependence of the external quantum efficiency $\eta_Q$ is resolved to that of several parameters of the semiconductor. 
Diode theory derives the external quantum efficiency to be 
\begin{eqnarray} 
\eta_Q = \epsilon\frac{L_h + L_e}{d}, 
\label{EQE}
\end{eqnarray}
where $\epsilon$ is emissivity of the diode, $d$ effective light penetration depth, and $L_{h/e}$ are diffusion length of the holes in the n-type semiconductor region and that of the electrons in the p-type semiconductor region, and they are assumed to be substantially longer than the width of the depletion layer (see Appendix).
With diffusion constant $D_{h/e}$ and lifetime $\tau_{h/e}$ for the minority carriers (holes and electrons), then the diffusion length is expressed as $L_h = \sqrt{D_h\tau_h}$ and $L_e = \sqrt{D_e\tau_e}$, respectively.

A possible explanation of the $\eta_Q-T_H$ curve for Diode A is a crossover between two minority carrier lifetimes.
It is known that the lifetime shows strong temperature dependence. 
The lifetime of each minority carrier is expressed by contribution of several interband transition mechanisms, such as $\frac{1}{\tau_{h/e}} = \frac{1}{\tau_\mathrm{Rad}}+\frac{1}{\tau_\mathrm{Auger}}+\frac{1}{\tau_\mathrm{SRH}}$. 
Here $\tau_\mathrm{Rad}$ is lifetime of the spontaneous interband transition with radiation of a photon.
The lifetimes $\tau_\mathrm{Auger}$ and $\tau_\mathrm{SRH}$ are caused by the non-radiative recombination mechanisms, known as Auger recombination and Shockley-Read-Hall (SRH) recombination, respectively.
At a temperature colder than $270\ \mathrm{K}$, the radiative recombination mainly occurs, and radiation lifetime $\tau_\mathrm{Rad}$ is shorter than others.
Therefore, $\tau_\mathrm{Rad}$ is dominant.
Above $270\ \mathrm{K}$, $\tau_\mathrm{Auger}$ or $\tau_\mathrm{SRH}$ decreases remarkably\cite{Beattie1958,Combescot1988PRB,Hoglund2014APL,Wen2015JAP}, $\eta_Q$ decreases as temperature increases.
For Diode B, radiative recombination is dominant in this measurement temperature range due to the larger $E_g$, and the decreasing of the $\tau_\mathrm{Auger}$ or $\tau_\mathrm{SRH}$ would be observed above $350\ \mathrm{K}$.

In Figure \ref{fig4} (a), the energy conversion efficiency $\eta_E$ of Diode A is lower than $10^{-4}\ \%$ at most, in contrast to high efficiency of $\eta_Q$. 
$\eta_E$ of Diode B is lower than that of Diode A, and shows a monotonic increase.
In eq. (\ref{chi}), we see that dominant factors for deciding $\eta_E$ are, $r_DS$, $\eta$, $y(x_g)$, and $T_H$
Increasing them to be larger values, the efficiency $\eta_E$ is going to be lager.
The diode or the hot bath temperature $T_H$ can be increased if we use some exhaust heat from industrial facilities. 
But for the concept of electrical power generation from terrestrial radiation, $T_H$ should be ambient temperature. 
The definite integral $y(x_g)$ increases as the bandgap energy $E_g$ decreases for a finite value of $T_H$, and in the limit for $E_g =0$, $y(0) = 2.404\cdots$. 
Thus, the contribution of $y(x_g)$ for improving efficiency is limited.
Similarly, the external quantum efficiency $\eta_Q$ is less than 1, contribution of $\eta_Q$ is also limited. 
The experimental results indicate that it is already enough high on the commercial devices.
The resistance of a p-n junction for a unit area $r_DS$, namely the inverse of conductivity per unit area, can be changed by many orders of magnitude by doping control. 
Therefore, $r_DS$ is a possible parameter to improve efficiency drastically.

Figure \ref{fig4}(b) shows the energy conversion efficiency $\eta_E$ as a function of $r_DS$ for $T_H = 300\ \mathrm{K}$ and for $\eta_Q = 50\ \%$ calculated with eq. (\ref{chi}). 
If $r_DS$ could be increased, the energy conversion efficiency $\eta_E$ would reach $100\ \%$.
This is because we assume $T_C =0\ \mathrm{K}$.

Increasing $r_DS$ is possible by increasing donor density $N_D$ or acceptor density $N_A$.
With the Shockley diode equation, $r_DS = k_BT_H/qJs$, where $J_s = qn_i^2 \left(\frac{1}{N_D}\sqrt{\frac{D_h}{\tau_h} }+\frac{1}{N_A}\sqrt{\frac{D_e}{\tau_e}}\right)$ is the reverse saturation current density, and $n_i^2 = N_CN_V \exp(-x_g)$ is square of the intrinsic carrier concentration.
The bandgap energy $E_g$ should be increased, however it conflicts to decreasing of $y(x_g)$.
The factor $y(x_g)/\exp(-x_g)$ has a maximum at $x_g \sim 2.52$, which corresponds to that the optimum bandgap energy is $E_g = 0.065\ \mathrm{eV}$ for energy conversion at $T_H=300\ \mathrm{K}$.

Figure \ref{fig4}(b) indicates that the output power per unit area $p_M/S$ can exceed the thermal radiation energy density $\sigma T_H^4$, if $r_DS$ is effectively increased. 
Upper limit for the open-circuit voltage $-v_{OC}$ must exist.
While $-v_{OC}$ is increased if $r_DS$ is increase, it must be not exceeding the difference of the quasi-Fermi level of minority carrier and that of majority carrier under the non-equilibrium condition.
The difference could be enhanced by carrier-density control, which is the same as controlling $r_DS$. 
A realistic limitation of $-v_{OC}$ is the reverse breakdown voltage of the diode.

Another question is what a fundamental trade-off relation between the output power and the energy conversion efficiency for the negative-illumination photovoltaic power generation is.
There must be a thermodynamical upper bound depending on the thermal radiation mechanism \cite{Buddhiraju2018PNAS,Shiraishi2016PRL,Pietzonka2018PRL}, however, exact theoretical expression of the efficiency is under investigation.
In our situation, to obtain maximum output power $P_M$ for an electrical load outside, the internal heating due to the shunt current flowing into the internal resistance $R_D$ is inevitable, and it is same with $P_M$.
If $P_M$ exceeds radiation energy flux $\sigma T_H^4$, the p-n junction is heated up. 
Thus, $P_M$ must be smaller than $\sigma T_H^4$, which means that the energy conversion efficiency must be smaller than 50\ \%.
As of now, the power generation with negative illumination is much smaller than these possible limitations. 
Therefore, the negative-illumination photovoltaic power generation leaves much to be improved.

\section{Conclusion}
In conclusion, we investigate the external quantum efficiency and the energy conversion efficiency of negative-illumination photovoltaic power generation of two mid-infrared diodes with different bandgap energy.
Using an experimental setup with a cold bath cooled by liquid nitrogen, inherent power generation from the diode device is experimentally observed.
We show a possible way to improve energy conversion efficiency is to decrease reverse saturation current density. 
Our experimental and theoretical approach will provides new insight for development of heat-electricity conversion devices based on narrow gap semiconductors with the negative-illumination photovoltaic effect, and also for the negative-illumination bulk-photovoltaic effect \cite{Lorenzi2020OpEx}, and near-field thermophotovoltaic energy conversion \cite{Mittapally2023PRApplied}.

\section*{Appendix}
\renewcommand{\theequation}{A.\arabic{equation}}
\setcounter{equation}{0}
The photocurrent $I_{ph}$ across a p-n junction of area $S$ is determined by the diffusion current at the junction interface, given by \cite{Grove1967text}
\begin{eqnarray} 
I_{ph} =-qS (G - R) (L_h + L_e),
\label{J_ph}
\end{eqnarray}
$L_h$ and $L_e$ are diffusion length of minority carriers (holes and electrons).
It is assumed that $L_h + L_e$ is substantially longer than the width of the depletion layer.
$G$ is the generation rate per unit volume and per unit time for electron hole pairs by absorption of a photon from outside, given by 
\begin{eqnarray} 
G(T_H,T_L) = \int_{E_g/h}^\infty \frac{\alpha(\nu,T_H)}{d(\nu,T_H)} \frac{2\pi\nu^2 }{c^2}\frac{1}{\exp(h\nu/k_BT_L)-1} d\nu.
\label{G}
\end{eqnarray}
$R$ is the radiative recombination rate by emitting a photon to outside, given by 
\begin{eqnarray} 
R(T_H) = \int_{E_g/h}^\infty \frac{\epsilon(\nu,T_H)}{d(\nu,T_H)} \frac{2\pi\nu^2 }{c^2}\frac{1}{\exp\{(h\nu-\mu)/k_BT_H\}-1} d\nu.
\label{R}
\end{eqnarray}
For simplification, $G$ and $R$ are assumed to be spatially homogeneous in both the p-type and n-type regions from the surface to light penetration depth $d$.
The diffusion length is assumed to be less than the light penetration depth. 
For the case $L_h + L_e > d$, diffusion length would be effectively limited by $d$, then $L_h + L_e$ should be replaced by $d$.
The chemical potential of photons $\mu$ is difference of quasi-Fermi levels \cite{Wurfel1982JPhysC}. 
Since $\mu$ is sufficiently smaller than the bandgap $E_g$ in our experimental situation, we assume that $\mu = 0$ in eq. (\ref{R}).
The absorptivity $\alpha(\nu,T_H)$ equals to $\epsilon(\nu,T_H)$ according to Kirchhoff's law of thermal radiation.
Neglecting $\nu$ dependence of $\epsilon(\nu,T_H)$ and $d(\nu,T_H)$ for simplification, equations (\ref{V_ph_fit}) and (\ref{EQE}) are derived.


\begin{acknowledgement}

The authors thank T. Hasegawa, Y. Tokizane and T. Tachizaki for fruitful discussions, and E. Sentoku and N. Horii for technical support.

\end{acknowledgement}

\newpage

\newpage

\newpage

\newpage

\newpage

\end{document}